\documentclass[twocolumn,preprintnumbers,amsmath,amssymb,superscriptaddress]{revtex4}
\usepackage{graphicx}
\usepackage{dcolumn}
\usepackage{bm}
\usepackage{natbib}

\def\ltsima{$\; \buildrel < \over \sim \;$}
\def\simlt{\lower.5ex\hbox{\ltsima}}
\def\gtsima{$\; \buildrel > \over \sim \;$}
\def\simgt{\lower.5ex\hbox{\gtsima}}

\def\[{\begin{equation}}
\def\]{\end{equation}}
\def\m@th{\mathsurround=0pt }
\def\eqalign#1{\null\,\vcenter{\openup1\jot \m@th
 \ialign{\strut\hfil$\displaystyle{##}$&$\displaystyle{{}##}$\hfil
 \crcr#1\crcr}}\,}

\begin{document}
\title{Locating the Baryon Acoustic Peak}

\author{Fergus Simpson}
\email{frgs@roe.ac.uk}
\author{John A. Peacock}
\author{Patrick Simon}
\affiliation{Institute for Astronomy, University of Edinburgh,
Royal Observatory, Blackford Hill, Edinburgh EH9 3HJ}

\date{\today}
\newcommand{\ud}{\mathrm{d}}
\newcommand{\und}{\underline}
\newcommand{\rbar}{y}

\begin{abstract}
Forthcoming photometric redshift surveys should provide an
accurate probe of the acoustic peak in the two-point galaxy
correlation function, in the form of angular clustering of
galaxies within a given shell in redshift space. We investigate
the form of the anticipated signal, quantifying the distortions
that arise due to projection effects, and in particular explore
the validity of applying the Limber approximation. A
single-integral prescription is presented, which provides an
alternative to Limber's equation, and produces a significantly
improved prediction in the regime of interest.

The position of the acoustic peak within the angular correlation
function relates to the angular diameter distance to the far side
of the redshift bin. Thicker redshift bins therefore shift
comoving features towards smaller angular scales. As a result, the
value of the photometric redshift error acquires a greater
significance, particularly at lower redshifts. In order to recover
the dark energy equation of state to a level of $1\%$, we find the
total redshift dispersion must be determined to within $\Delta
\sigma_z \lesssim 10^{-3}$, which may prove challenging to achieve
in practice.

\end{abstract}
\maketitle

\section{Introduction}

The propagation distance of sound waves prior to decoupling is a
convenient reference scale embedded in the distribution of
galaxies. Lately these Baryon Acoustic Oscillations (BAO) have
received particular attention due to their potential to measure
the dark energy equation of state, $w$
\cite{2003ApJ...594..665B,2003ApJ...598..720S,2005ApJ...633..560E,2005MNRAS.362..505C}.
To attain competitive constraints on $w$, the angular scale of the
BAO must be determined at the sub-percent level. In addition,
robust models are required for any phenomena capable of inducing
systematics of a comparable magnitude. In the near future, large
photometric redshift surveys such as Pan-STARRS
\footnote{http://pan-starrs.ifa.hawaii.edu/public/} and DES
\footnote{http://www.darkenergysurvey.org/} will cover a
substantial proportion of the sky, providing a vast map of large
scale structure in the universe. However, the precision and
reliability of the photometric redshifts will significantly
influence the dark energy constraints. With a typical rms redshift
error of $\sigma_z \sim 0.05$ the longitudinal modes of the power
spectrum are exponentially damped, leaving only those which are
closely aligned to the plane of the sky. Therefore BAOs from
photometric redshift surveys are predominantly sensitive to the
angular diameter distance $D_A$. This may also be interpreted as
an effective reduction in volume given by approximately
$12(\sigma_z/(1 + z)/0.03)$ \cite{2005MNRAS.363.1329B}, although
Cai et al. \cite{2008arXiv0810.2300C} reach a more optimistic
conclusion, suggesting only a fivefold reduction in volume.

\begin{figure*}
\includegraphics[width=120mm]{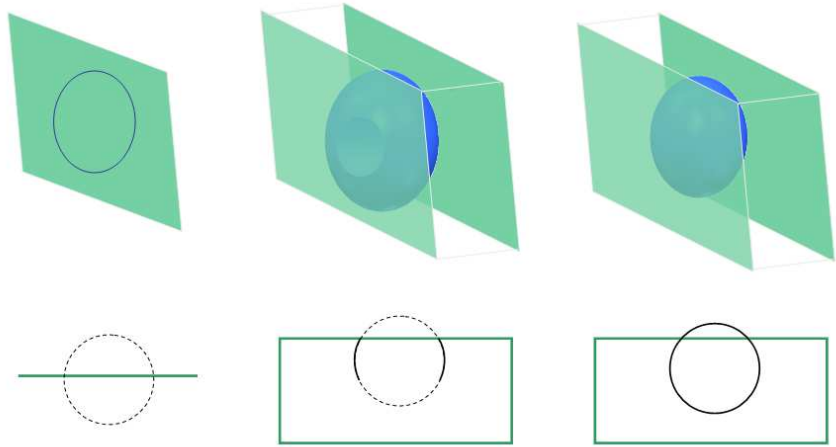}
\caption{A representation of the observable pairings in the case
of (i) a thin redshift slice (ii) a thick redshift slice with
truncation, as described in (\ref{sec:finite}) and (iii) a thick
redshift slice without truncation, as provided by the Limber
approximation (\ref{eq:limber}). Each point on the shell's surface
-- along with its corresponding antipode -- represents a potential
galaxy pair. } \label{fig:trunc}
\end{figure*}

The characteristic signature for a particular dark energy model
resides in the apparent redshift variation of the angular acoustic
scale. In order to study this behaviour, the galaxy population may
be divided into thin redshift slices, limited in thickness to be
$\gtrsim \sigma_z$. We explore the smoothing effect which arises
when projecting from three dimensions down to two. Locating the
acoustic signature in Fourier space and real space -- via the
power spectrum and correlation function -- are each susceptible to
different forms of systematic errors. Finding both in agreement
would therefore strengthen our confidence in any subsequent
constraints on cosmological parameters. For this study we focus on
the acoustic peak in the angular correlation function, paying
particular attention to the influence of projection effects, and
analyse the shift in peak position as a function of bin width. The
Limber approximation, recently examined by Simon
\cite{2006astro.ph..9165S}, acts as a starting point from which we
develop an alternative analytic description of the transformation
from the spatial to angular correlation function. Recently LoVerde
\& Afshordi \cite{2008arXiv0809.5112L} express the angular power
spectrum as a power series, within which the Limber approximation
appears as the $0^{th}$ order term. Our approach is somewhat
different, and we obtain a single-integral approximation for the
angular correlation function, which provides a good description
 for galaxy redshift distributions of practical interest.

\section{Fiducial Cosmology}

To evaluate the spatial correlation function we begin with the
linear matter power spectrum output from CAMB
\cite{2000ApJ...538..473L}, and adopt the default cosmological
parameters $\Omega_m = 0.27; \Omega_\Lambda = 0.73; \Omega_b =
0.046; n_s = 0.96; \sigma_8=0.79$. Throughout this work we assume
a flat universe.

Conventional means of modelling non-linear structure formation,
such as the HALOFIT prescription \cite{2003MNRAS.341.1311S} are
not well suited to an oscillatory $P(k)$ since they focus on
corrections in the highly non-linear regime, rather than the
quasilinear region where the BAO signal resides. Recently
Eisenstein et al. \cite{2005ApJ...633..560E} offered a simple
approach to modelling the degradation of the acoustic peak. By
considering the rms radial displacement for a pair of galaxies
separated by $100 \, h^{-1} \mathrm{Mpc}$, the damping is given by

\[
P(k)=P_{\mathrm{linear}}(k) \, e^{-\frac{k^2 \sigma^2}{2}} ,
\]

\[
\sigma = s_0 D ,
\]

\noindent where $D \propto \delta(z)/\delta(0)$ is the growth
function (normalised such that $D=a$ at early times), and $s_0$ is
taken to be $10.9 \, h^{-1} \mathrm{Mpc}$, rescaled from the $12.4
\, h^{-1} \mathrm{Mpc}$ quoted in \cite{2005ApJ...633..560E} to
match our lower value of $\sigma_8$. The real-space correlation
function at a given redshift is then obtained by

\[
\xi(r) = \frac{1}{2 \pi^2} \int P(k) \, \frac{k}{r} \, \sin (kr)
\, \, \ud k .
\]

\noindent However in redshift space the power is anisotropic, so
before proceeding further we first explore the extent to which
redshift distortions may impact our analysis. The anisotropic
correlation function $\xi(r_\perp, r_\parallel)$, where $r_\perp$
and $r_\parallel$ are the transverse and radial parts of
$\und{r}$, is related to the power spectrum via

\[ \eqalign{
& \xi(r_\perp, r_\parallel) = \frac{1}{(2 \pi)^3} \int P(\und{k})
\, e^{i \und{k} \cdot \und{r}} \, \ud^3 k \cr  & = \frac{1}{(2
\pi)^2} \int_{-1}^1 \int_0^\infty P(k,\mu) \, J_0\left(k r_\perp
\sqrt{1-\mu^2} \right) e^{i \mu k r_\parallel} \, k^2 \, \ud k \,
\ud \mu }
\]

\noindent where $\mu k$ is the radial part of $\und{k}$. Thus e.g.
$\xi(r_\perp,0)$ is not in general the same as the real-space
correlation function. Incorporating a simple model of the Kaiser
 and Fingers of God effects leaves us with

\[\eqalign{
\xi(r_\perp,0) = & \frac{1}{(2 \pi)^2} \int_{-1}^1 \int_0^\infty
P(k) \, J_0 \left(k r_\perp \sqrt{1-\mu^2}\right) \times \cr & (1
+ \beta \mu^2)^2 \, e^{-\mu^2 k^2 (\sigma_z^2+\sigma_v^2) } \, k^2
\, \ud k \, \ud \mu }
\]

\noindent where $\sigma_v$ denotes the one-dimensional rms
single-point velocity dispersion. The strong damping of radial
modes ensures that for all but the smallest values of $k$,
contributions from the Kaiser effect are restricted to small
values of $\mu$. This resembles a slow-varying scale-dependent
bias in the angle-averaged power spectrum, which is therefore
unproblematic provided we adopt an empirical ``wiggles only"
fitting function.

In practice we will want to consider angular clustering in shells
of width comparable to the photo-z error, $\sigma_z \simeq 0.05$.
This scale is large by comparison with those of peculiar
velocities, so we shall ignore the above complication and treat
the situation as though photo-z's are a noisy measure of the true
radius, even though this is not strictly true. We shall thus tend
to use the term `redshift shell' to mean `radial shell'.

\begin{figure*}
\includegraphics[width=80mm]{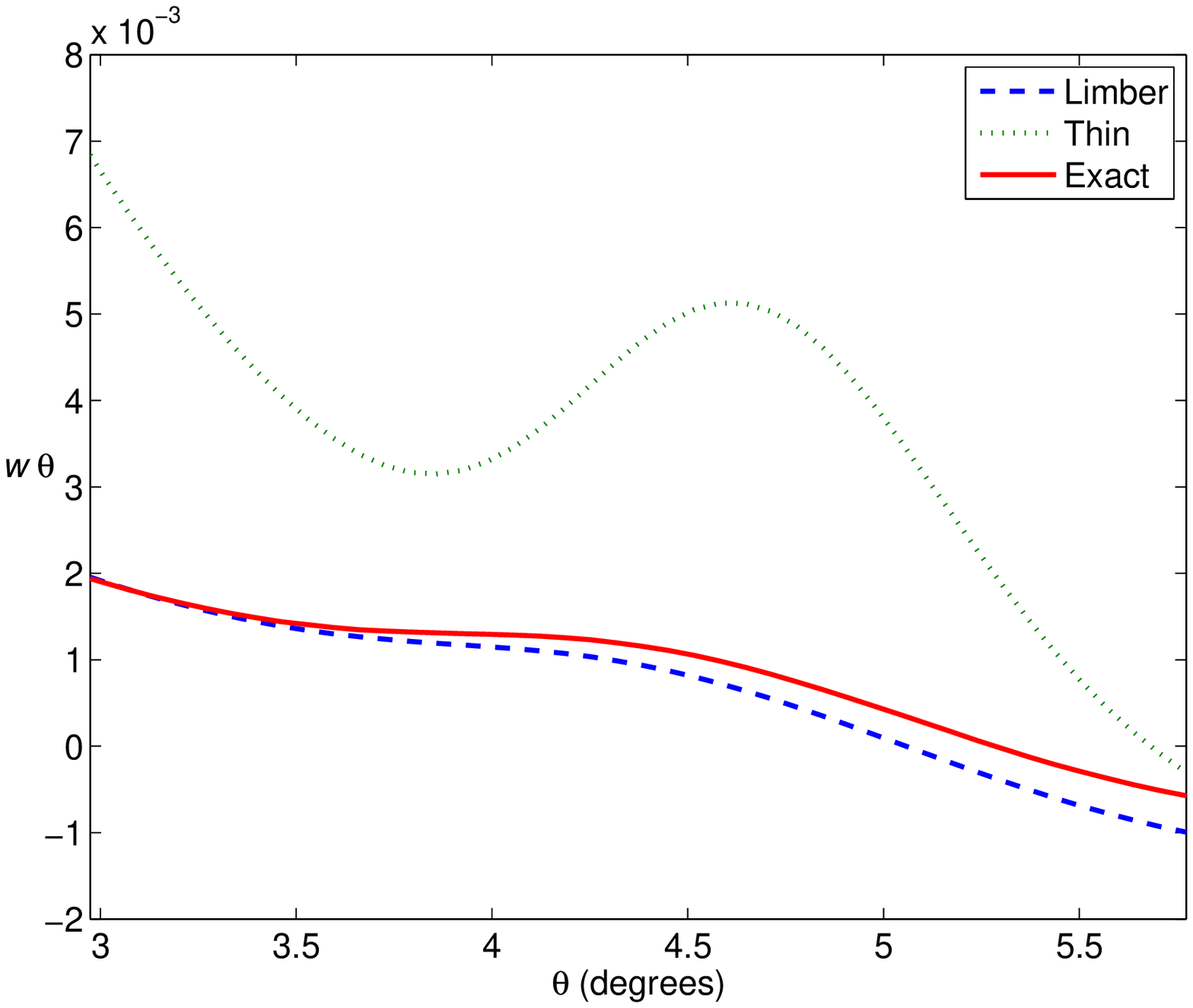}
\includegraphics[width=80mm]{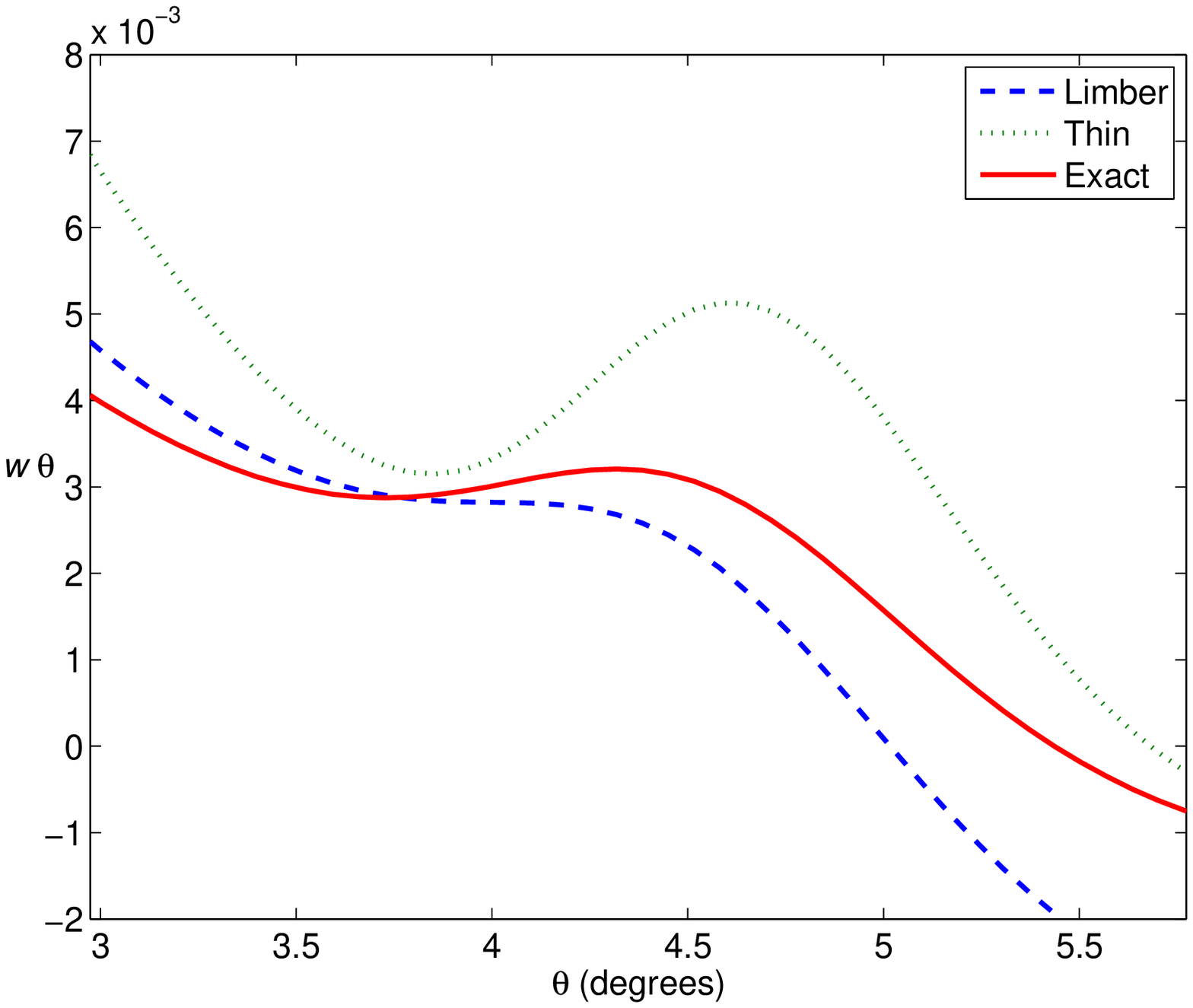}
\caption{The angular correlation function (solid) for a top-hat
galaxy redshift distribution evaluated at $z=0.5$, compared to the
Limber approximation (dashed), and thin-layer approximation
(dotted). The bin widths are scaled such that $\Delta z=0.1 (1+z)$
(left) and $\Delta z=0.04 (1+z)$ (right).} \label{fig:peaks}
%
\includegraphics[width=80mm]{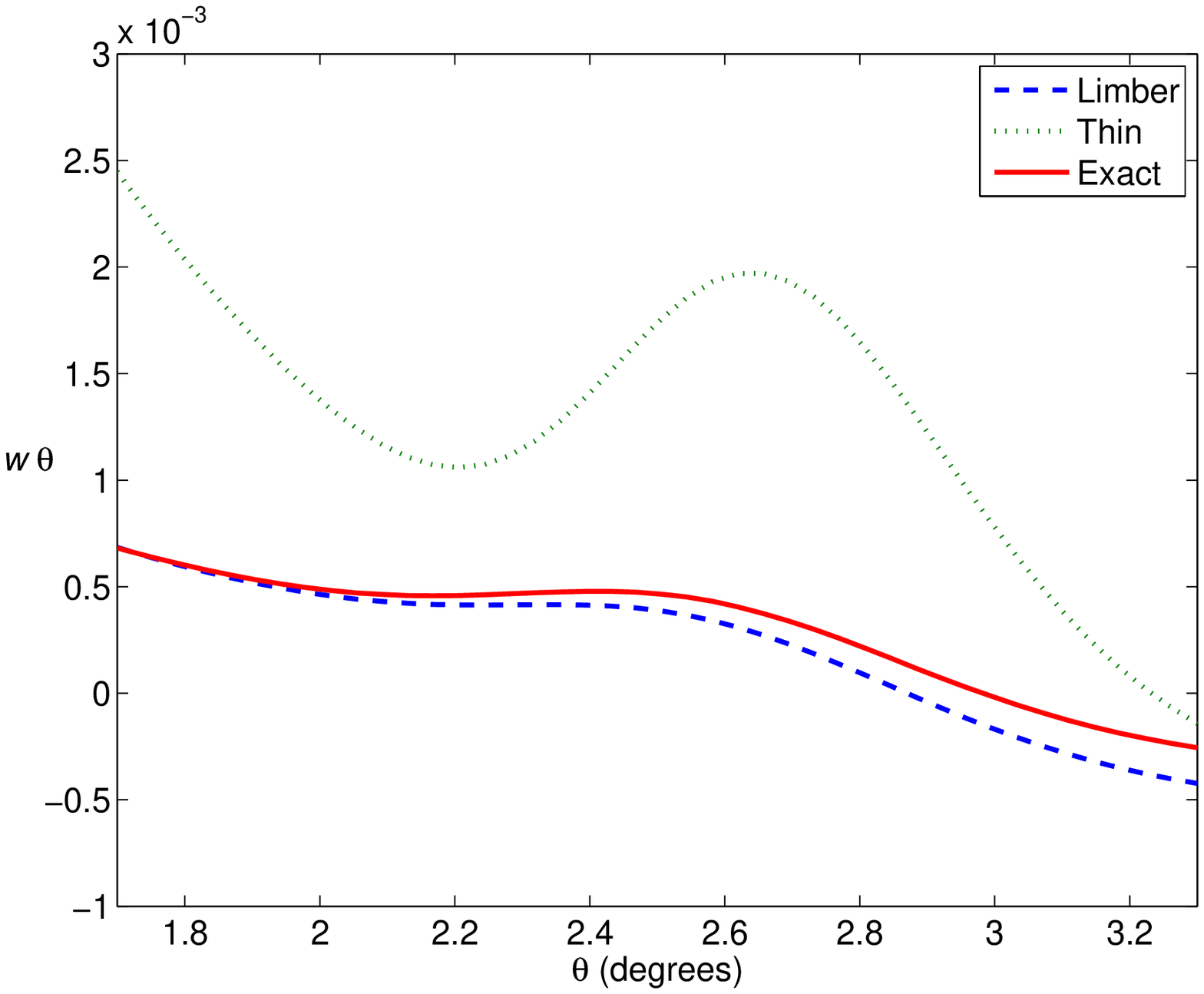}
\includegraphics[width=80mm]{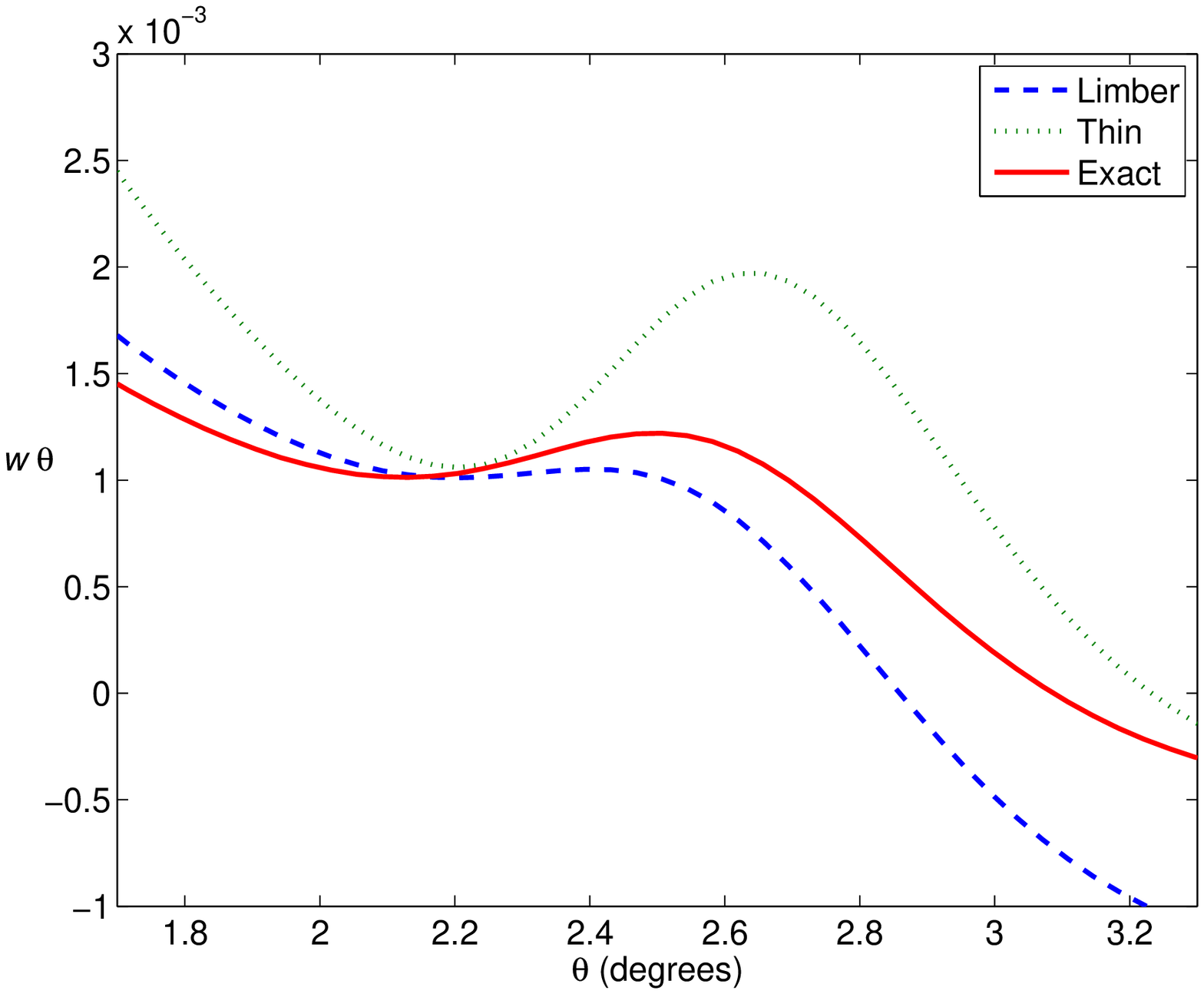}
\caption{ The same plots as in Fig.\ref{fig:peaks}, but with the
bin located at $z=1$.} \label{fig:peaks2}
%
\includegraphics[width=80mm]{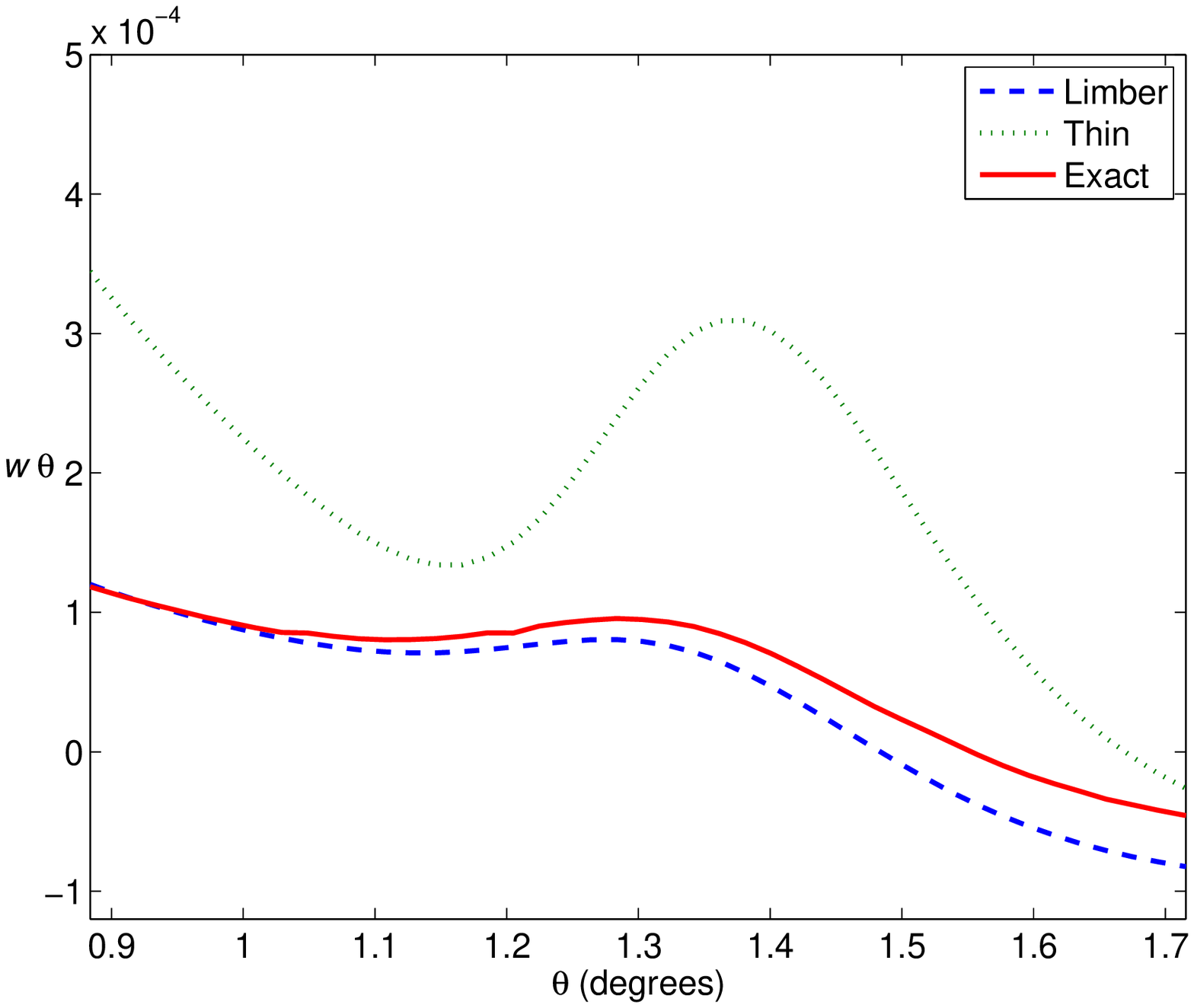}
\includegraphics[width=80mm]{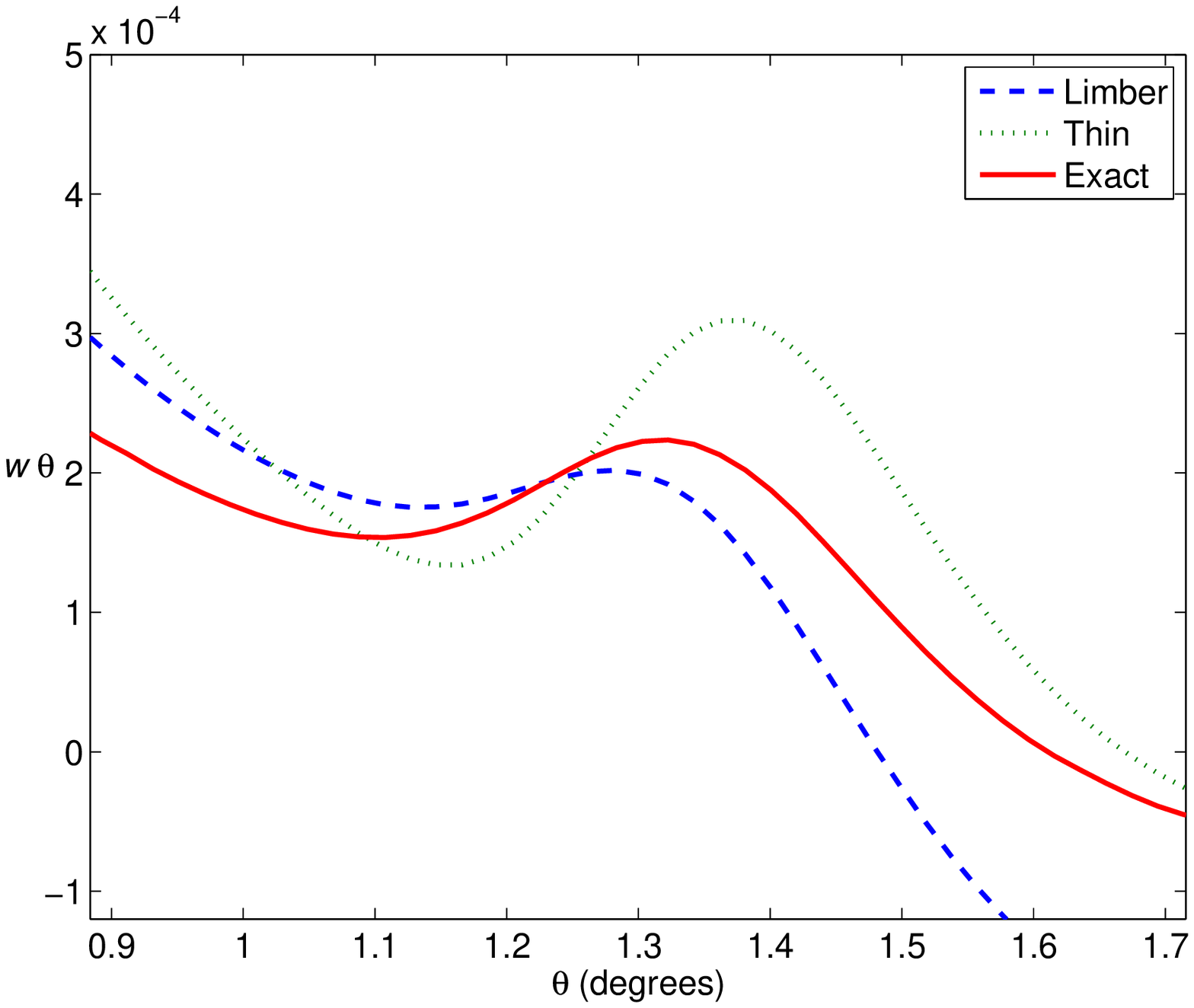}
\caption{ The same plots as in Fig.\ref{fig:peaks}, but with the
bin located at $z=3$.} \label{fig:peaks3}
\end{figure*}

\section{Projection Effects}

To begin, we consider an infinitesimally thin radial shell,
 so clustering may only be viewed perpendicular to the line of sight.
In this case the angular correlation function simply resembles a
rescaled version of the spatial correlation function, $w_t
(\theta) \simeq \xi(\chi_s \theta)$, where $\chi_s$ is the
(comoving) radial distance to the bin. This is known as the
thin-layer approximation. Any given part of the spatial
correlation function could therefore be represented as a circle on
the sky, as illustrated in the left hand panel of
Fig.~\ref{fig:trunc}.

When considering a thicker redshift bin, with a uniform selection
function, two distinct phenomena influence our observation. The
first, as represented by a truncated sphere in the centre panel of
Fig.~\ref{fig:trunc}, is that galaxy pairs with significant
line-of-sight displacements begin to contribute to the signal.
Therefore any given large-scale feature from $\xi(r)$ begins to
influence the smaller angular scales of $w(\theta)$. The extent of
this is governed by the ratio of the clustering scale to the bin
thickness. The second effect arises since the truncated sphere
subtends a different angle depending on its location within the
bin. This effect gains prominence when the bin thickness is a
significant proportion of the radial distance to the bin. This can
often be alleviated by adopting a projected correlation function
rather than an angular correlation function, but the presence of
photometric redshift error proves to be a limiting factor.

For the case of a redshift bin much thicker than the clustering
scale, as seen in the right hand panel of Fig.~\ref{fig:trunc},
our model is simplified since any truncation can be neglected and
only the projection effect persists. Therefore any galaxy pair
whose \emph{centre} lies within the bin is assumed to contribute
to the signal. This simplification leads to the erroneous
inclusion of pairs which extend beyond the edge of the bin. Indeed
this is the situation described by the Limber approximation, which
provides us with a good starting point for the purposes of
studying the acoustic peak. The clustering scale of interest is
around $150 \, \mathrm{Mpc}$, whilst photometric redshift
precision ensures the effective bin thickness is typically double
this.

\subsection{Limber Shift} \label{sec:limb}

The angular correlation function is  evaluated numerically using
(23) from Simon \cite{2006astro.ph..9165S}

\[
w (\theta) = \frac{2}{1+\cos \theta} \iint \frac{\ud n}{\ud \rbar}
\Big |_{\rbar-\frac{x}{2}} \frac{\ud n}{\ud \rbar} \Big
|_{\rbar+\frac{x}{2}} \xi(\rbar,R) \frac{R}{x} \, \ud R \, \ud
\rbar , \label{eq:exact}
\]

\noindent where the variables $R$, $\rbar$, and $x$ denote the
galaxy pair separation, their mean distance from the observer, and
their line-of-sight separation, as given by

\[
x \equiv \sqrt{\frac{2R^2-4\rbar^2(1- \cos \theta)}{1+\cos
\theta}}, \label{eq:delta}
\]

%

\noindent The evolution of $\xi$ is represented by its first
argument, whilst the second provides the galaxy pair separation
$R$. This formulation assumes that only linear evolution of $\xi$
occurs over the distance $x$, and no significant change in
position of the nearer galaxy arises during the light-travel time
between the pair.

Projection effects lead to the acoustic peak appearing at a
smaller angular scale than one might na\"{\i}vely expect. Here we
assess whether a simplified form of (\ref{eq:exact}) may be used
to provide an estimate of the acoustic peak location. The Limber
approximation is given by

\[\label{eq:limber}
w (\theta) = \int_0^\infty \! \int_{-\infty}^{\infty} \left(
\frac{\ud n}{\ud \rbar} \right)^2 \xi \left(\rbar,
\sqrt{x^2+\rbar^2 \theta^2} \right) \, \ud x \, \ud \rbar  .
\]

This simplification utilises the flat-sky approximation, which
holds well even on the relatively large angular scales under
consideration. At a separation of $5^\circ$ the approximation
$(1+\cos \theta) \sim 2$ provides an overestimation of $0.2\%$,
and this propagates to $R$ with an error of less than $0.1\%$.
This is more prominent for the longitudinal modes, to which we are
least sensitive.

Figs.~\ref{fig:peaks}-\ref{fig:peaks3} illustrate the shift in the
acoustic peak as predicted by the Limber equation. Note how the
acoustic peak is almost entirely erased when the bin width is
$\Delta z=0.1(1+z)$, highlighting the need for accurate
photometric redshifts. The increasing bin thickness, chosen to
mimic the degradation of the photometric redshifts, is
counteracted by the reduced value of $\ud \chi / \ud z$ at higher
redshift, leading to little change in the projection effect across
the range of redshifts $0.5 < z < 3$. Ordinarily the Limber
approximation overestimates the signal, however an underestimate
arises here due to the influence of the correlation function's
negative regime. The Limber approximation clearly offers a more
appropriate estimation than that of the thin layer. Its precision
varies according to the nature of the chosen fitting algorithm,
and improves with broader bins. We find the relevant angular scale
is underestimated, with the magnitude of the discrepancy typically
ranging from $2\%$ to $5\%$, substantially exceeding the level of
accuracy required for dark energy studies.

\section{Shift Estimation}

As we have seen, adopting the Limber approximation results in a
good but insufficiently accurate estimation for the positioning of
the acoustic peak. Here we attempt to gain a quantitative
understanding of the shift, and interpret this as a smearing of
$w_t(\theta)$, the correlation function from an infinitesimal bin.

\subsection{Limber Equation}

We begin by transforming the Limber equation to plane polar
coordinates

\[\label{eq:limberpolar1}
w (\theta) = \frac{2}{\theta}   \int_{0}^{\pi /2} \int_0^\infty
\left(\frac{\ud n}{\ud \rbar}\right)^2 \xi(r) \, r \, \ud r \, \ud
\theta' ,
\]

\noindent where we substitute $x=r \sin \theta'$, $\rbar \theta=r
\cos \theta'$.

For the moment we neglect the redshift evolution in the
correlation function, and assume a top-hat galaxy distribution,
simplifying our equation by assuring $\ud n / \ud y=(\Delta
\chi)^{-1}$ within the top-hat, where $\Delta \chi$ corresponds to
the physical depth of the redshift bin.  Later we shall see that
our results are largely insensitive to the functional form of the
redshift bin, and the impact of an evolving $\xi(\chi)$ across
reasonably narrow redshift bins is minimal.

These simplifications leave us with an angular correlation
function given by

\[\label{eq:limberpolar2}
w (\theta) =  \frac{2}{\theta \Delta \chi^2}
\int_{\theta'_1}^{\theta'_2} \int_0^\infty
 \xi(r) \, r \, \ud r \, \ud \theta'  ,
\]

\noindent and it is instructive to study the case where the
correlation function consists of a delta function at an arbitrary
location $r_0$ :

\[\label{eq:deltafn}
 \xi(r) = \delta(r_0 - r) ,
\]

\[\label{eq:limberpolar3}
w (\theta) =  \frac{2 r_0}{\theta \Delta \chi^2}
\int_{\theta'_1}^{\theta'_2}  \, \ud \theta' ,
\]

\noindent where the lower integration limit is given by

\[\label{eq:theta1}
\theta'_1 = \arccos\left(\frac{\chi_1 \theta}{r_0}\right) ,
\]

\[\label{eq:chi1}
\chi_1 = \chi_s + \frac{\Delta \chi}{2} ,
\]

\noindent  and similarly for $\theta'_2$:

\[\label{eq:theta2}
\theta'_2 = \arccos\left(\frac{\chi_2 \theta}{r_0}\right) ,
\]

\[\label{eq:chi2}
\chi_2 = \chi_s - \frac{\Delta \chi}{2} ,
\]

\noindent where we denote the upper and lower boundaries of the
galaxy distribution by $\chi_1$ and $\chi_2$ respectively, and
define $\arccos(x)=0$ for $x>1$. This leaves us with

\[\label{eq:lim}
w_{s}(\theta) = \frac{2 r_0}{\theta \Delta \chi^2} \left[\arccos
\left( \frac{\chi_2\theta}{r_0}\right)-\arccos\left(
\frac{\chi_1\theta}{r_0}\right)\right] .
\]

\noindent Substituting $r_0 = \chi_s \theta_0$ leaves us with the
Limber smear function

\[\label{eq:limsmear}
w_s(\theta) = \frac{2 \chi_s^2 \theta_0}{\theta \Delta \chi^2}
\left[\arccos \left(\frac{\chi_2\theta}{\chi_s
\theta_0}\right)-\arccos\left(\frac{\chi_1\theta}{\chi_s
\theta_0}\right)\right] .
\]

\begin{figure}
\includegraphics[width=80mm]{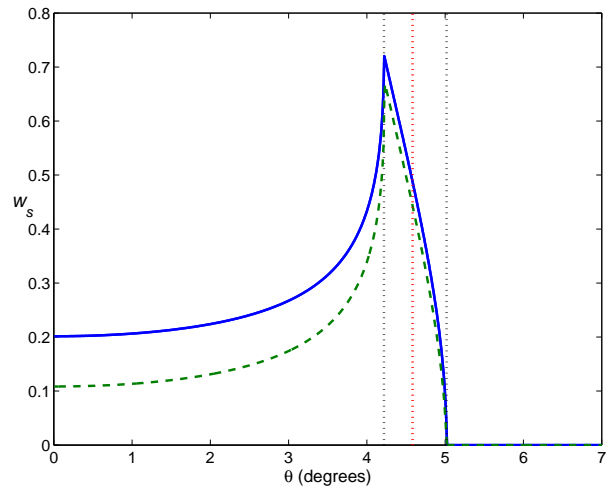}
\caption{The solid line represents the smear function derived from
the Limber approximation, $w_s$ as defined in (\ref{eq:limsmear}),
and evaluated at $z=0.5$ with $\Delta z=0.1$. The dashed line is
the exact expression, illustrating the modification introduced by
accounting for the finite bin width. On smaller angular scales,
corresponding to a large line-of-sight separation, the pairs lost
from truncation become particularly apparent. Of the three
vertical dotted lines, the central one represents the original
$\delta$-function, while the left and right lines denote the
angular scales associated with the far and near bin edges
respectively. } \label{fig:smear}
\end{figure}

The functional form of this is illustrated in
Fig.~\ref{fig:smear}, and this highlights the mechanism by which
the peak position is shifted to smaller angular scales, typically
by around $10\%$. The projected signal peaks at $\theta_p$, the
angle corresponding to the far edge of the redshift bin, as given
by

\[\label{eq:thetashift}
\theta_p = \frac{\chi_s}{\chi_1} \theta_0.
\]

\noindent  At greater angular scales contributions are rapidly
lost, starting from those at the far edge of the bin, with the
signal reaching zero at the angular scale corresponding to the
near edge of the bin. Conversely, smaller angular scales gradually
receive fewer contributions since they require the pair
orientation to be increasingly aligned with the line of sight.
These are the pairs which prove particularly susceptible to
overestimation by the Limber equation, since they readily protrude
beyond the bin boundary.

Now returning to (\ref{eq:limsmear}), and extending our analysis
to a general correlation function, the parameter $\theta_0$
becomes a variable over which we integrate

\[\label{eq:limsmear2}
w_{\rm{L}}(\theta) = \int_0^\infty w_s(\theta,\theta_0) \,
\xi(\chi_s \theta_0) \, \ud \theta_0.
\]

\subsection{Correcting for Bin Width} \label{sec:finite}

Here we modify the limits of our integration in order to account
for the finite bin width. This aims to improve on the Limber
approximation's estimate for the location of the acoustic peak.

Towards the bin edges the spherical shell assumption breaks down
-- a galaxy cannot pair up with another galaxy which lies beyond
the bin boundary. This truncation is particularly problematic on
scales which are large relative to the bin width (see
Fig.~\ref{fig:trunc}). By considering the point at which one of
the galaxy pair extends beyond the bin boundary, we shrink the
integration range accordingly, with our new limits given by

\[\label{eq:thetatr1a}
\theta'_1 = \arccos\left[\frac{(\chi_1 - \delta \chi_1) \,
\theta}{r_0}\right] ,
\]

\noindent where the distance from the bin edge $\delta \chi_1$ is

\[\label{eq:thetatr1b}
\delta \chi_1 = \frac{\chi_s \theta_0}{2} \cos \left(
\frac{\theta}{2} -
 \phi_1 \right)  ,
\]

\[\label{eq:thetatr1c}
\sin \phi_1= \frac{\chi_1 \theta}{\chi_s \theta_0} \cos \left(
\frac{\theta}{2} \right)  ,
\]

similarly

\[\label{eq:thetatr2a}
\theta'_2 = \arccos\left[\frac{(\chi_2 + \delta \chi_2) \,
\theta}{r_0}\right] ,
\]

\[\label{eq:thetatr2b}
\delta \chi_2 = \frac{\chi_s \theta_0}{2} \cos \left(
\frac{\theta}{2} + \phi_2 \right)  ,
\]

\[\label{eq:thetatr2c}
\sin \phi_2= \frac{\chi_2 \theta}{\chi_s \theta_0} \cos \left(
\frac{\theta}{2} \right)  .
\]

\noindent The angular correlation function is then recovered by

\[\label{eq:correct2}
w(\theta) = \int_0^\infty   w_c(\theta,\theta_0)  \, \xi(\chi_s
\theta_0) \, \ud \theta_0  ,
\]

\[\label{eq:correct2b}
w_c(\theta,\theta_0) = \frac{2 \chi_s^2 \theta_0}{\theta \Delta
\chi^2} \left(\theta'_1 - \theta'_2\right) .
\]

\noindent This expression is exact within the context of a uniform
galaxy distribution and flat sky approximation.

The influence of this adjustment is illustrated by the dashed line
in Fig.\ref{fig:smear}. The Limber equation generally
overestimates the amplitude of the correlation function due to
over-counting the number of pairs, and this truncated integral
provides the necessary correction.  Upon evaluating the integral,
we recover the solid and dashed lines originally shown in
Fig.\ref{fig:peaks}.

\section{The galaxy distribution}

Up until now we have considered a uniform galaxy distribution.
Provided $\Delta \chi/\chi$ is small, the functional form of
$n(z)$ has a limited impact on the appearance of the correlation
function. We find that the $w(\theta)$ produced by a Gaussian
selection function may be closely mimicked by simply selecting a
top-hat selection function, of width $\Delta z$ chosen with a
matching rms, given by

\[\label{eq:nz}
\Delta z = \sqrt{12} \sigma(n(z)).
\]

\begin{figure}[t]
\includegraphics[width=80mm]{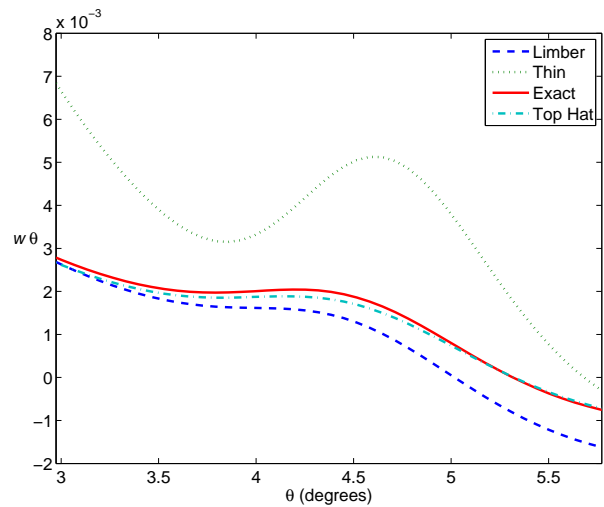}
\caption{Here we take the angular correlation function
corresponding to a Gaussian redshift distribution at $z=0.5$ with
$\sigma_z=0.02(1+z)$, and superpose that of a uniform galaxy
distribution (dot-dash), scaled to provide a matching
 rms.} \label{fig:gauss}
\end{figure}

Fig.~\ref{fig:gauss} compares three estimates of the angular
correlation function for a Gaussian distribution. We find that
this top-hat approximation compares favourably to the conventional
Limber and thin layer approaches, producing a more faithful
reproduction of the bump's shape and magnitude. Yet for higher
precision and larger values of $\Delta z/z$, inevitably one must
return to the exact form given by (\ref{eq:exact}).

\begin{figure}[t]
\includegraphics[width=80mm]{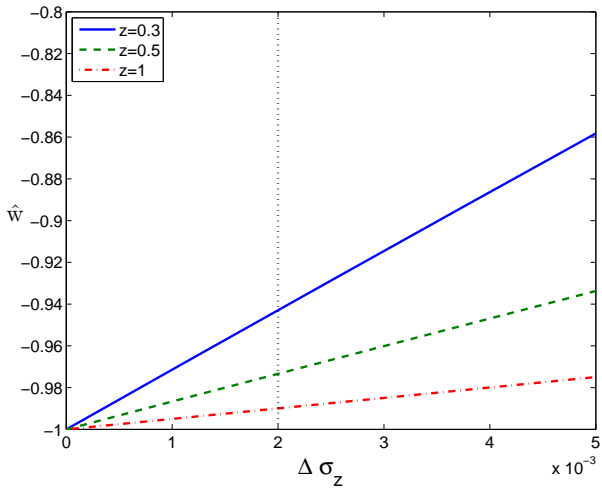}
\caption{ Overestimating the photometric redshift error may lead
to a substantial overestimate of the dark energy equation of
state, particularly at lower redshifts. The vertical line denotes
the achievable limit in the near-term future, as given by Newman
\cite{2008ApJ...684...88N}.} \label{fig:shifting}
\end{figure}

\section{Calibration Requirements}

In this section we clarify what constraints on the photometric
redshift errors are needed in order to avoid significant
degradation or biasing of the dark energy constraints from future
surveys.

\subsection{Photometric Redshift Error}

When projecting from the 3D correlation function to the
two-dimensional angular correlation function, not only does the
smoothing process dampen the appearance of the peak, but its
angular position shifts in relation to the thickness of the
redshift bin. This modified peak location, previously illustrated
in Figure \ref{fig:smear}, is associated with the angular diameter
distance at the \emph{far edge} of the redshift bin. It is via
this mechanism that the magnitude of the photometric redshift
error controls the peak position. Any uncertainty in $\sigma_z$
therefore influences our estimate of the dark energy equation of
state.

To quantify this effect we select a top-hat of width $\Delta
z_p=0.1$ in photometric redshift space, which is then convolved
with a Gaussian of standard deviation $\sigma_z$ to simulate the
true redshift distribution. As demonstrated in the previous
section, the resulting angular correlation function is readily
approximated by simply adopting a pure top-hat selection function
of width

\[
\Delta z =\sqrt{\Delta z_p ^2 + 12 \sigma_z^2},
\]

\noindent and the redshift of interest is $z_1$, the far edge of
the bin

\[
z_1  =  \bar{z} + \frac{\Delta z}{2} . \label{eq:z1}
\]

It is therefore crucial to acquire accurate and unbiased estimates
of both the mean redshift $\bar{z}$ and the photometric error
$\sigma_z$. Note that from (\ref{eq:z1}) we find

\[
\eqalign{
 \frac{\ud z_1}{\ud \sigma_z} & =  \frac{6 \sigma_z}{\Delta z} \cr
   } \label{eq:dz1} ,
\]

\noindent thus if we take $\Delta z_p = 2 \sigma_z$, this leaves
our measurement $50\%$ more sensitive to the value of $\sigma_z$
compared to the mean redshift $\bar{z}$. A selection function much
wider than $2 \sigma_z$ is undesirable as it leads to excessive
damping of the peak, and leaves us susceptible to the redshift
evolution  of $\sigma_z$.

To proceed, we assume the mean redshift is perfectly known, and
explore the impact of $\sigma_z$ on our estimation of the dark
energy equation of state.

\subsection{Dark Energy Constraints}

The process of fitting the acoustic peak to the data is not
straightforward. A strict fitting regime corresponding to the
anticipated shape of $\xi(r)$ from simulations would leave us
vulnerable to systematic errors. Meanwhile the adoption of a fitting   
function with many degrees of freedom would substantially degrade
the precision of our measurement. To avoid introducing any
dependence on the fitting method into our results, we simply
consider the angular shift induced on a correlation function of
the form $\xi(r) = \delta(r_0-r)$, where $r_0 = 110 \, h^{-1}
\mathrm{Mpc }$.

In Figure \ref{fig:shifting} we evaluate the impact of an
uncertainty in $\sigma_z$ at redshifts $z=0.3,0.5,1$, with fixed
bin width $\Delta z_p=0.1$ and $\sigma_z=0.05(1+z)$. We
incorporate an error $\Delta \sigma_z$ defined such that

\[
\Delta z =  \sqrt{\Delta z_p ^2 + 12 (\sigma_z + \Delta
\sigma_z)^2},
\]

Our estimate of the equation of state, $\hat{w}$, is then
perturbed by an amount relating to $\ud D_A / \ud z_1$, thus
leading to the steeper slopes at lower redshift. Provided the
estimator is unbiased, averaging over several bins could improve
the situation. Conversely, bypassing the CMB by taking the ratios
of angular scales in different bins would further aggravate the
problem.


\subsection{Measuring $\sigma_z$}

The conventional method for estimating the photometric redshift
error is to use a calibration set of spectroscopic redshifts
covering a suitably representative sample of galaxies. Newman
\cite{2008ApJ...684...88N} provided an estimate of $\Delta
\sigma_z$ attainable from spectroscopic calibration samples,
combined with information from their angular clustering, and found
$\Delta \sigma_z \sim 2 \times 10^{-3}$. This includes a modelling
of the outlier population as a second Gaussian with a
significantly larger standard deviation.

Zhan \& Knox \cite{2006ApJ...644..663Z} consider using the
exponential damping of the longitudinal modes of the galaxy power
spectrum to determine the ratio $\sigma_z/H$, where $H$ is the
Hubble parameter at the redshift of interest. Applying a prior on
$H$ will then provide an estimate of $\sigma_z$. For example, for
an LSST-like survey, they find a $1\%$ prior on $H$ leads to an
uncertainty on $\sigma_z$ of $1\%$. Ma et al.
\cite{2006ApJ...636...21M} explore the impact of photometric
redshift uncertainties on weak lensing surveys, which appear to
have photometric calibration requirements at a similar level to
that of the BAO.


The non-Gaussianity of photometric redshift errors, along with
their strong variation with redshift (see Fig 8 of
\cite{2008arXiv0810.2300C}), ensures that we require more degrees
of freedom than simply $\sigma_z$, which further heightens the
difficulty of reaching the required level of precision. Ma \&
Bernstein \cite{2008ApJ...682...39M} estimate this may be
compensated for by enlarging the spectroscopic calibration sample
by a factor of five.



\section{The Projected Correlation Function}

Here we justify the choice of an angular correlation function,
used throughout this work, as opposed to the projected correlation
function. This is defined such that

\begin{equation} \label{eq:projected}
w_p(r_\perp) \equiv 2 \int_{r_\perp}^\infty \xi(r) \,
\frac{r}{\sqrt{r^2 - r^2_\perp}} \, \ud r .
\end{equation}

\noindent To evaluate this quantity we require the precise
location of the galaxy in the bin, but this is unknown since
$\sigma_z \sim \Delta z$. As a result, adopting this approach
aggravates the smoothing of the correlation function. To
illustrate this we consider the range of transverse separations
which are observed by a particular value of $w(\theta)$ and
$w_p(\theta \chi_s)$. For the angular correlation function this is
simply given by the limits $\chi_1 \theta$ and $\chi_2 \theta$.
However with the projected correlation function this window is
extended to $(\theta \chi_s \chi_1 /\chi_2)$ and $(\theta \chi_s
\chi_2 /\chi_1)$ respectively. This occurs when pairs thought to
be located at $\chi_1$ are actually at $\chi_2$, and vice versa.

Conversely when studying a bin with $\sigma_z < \Delta z$ the
projected correlation function is the more natural option, and
will be subject to projection effects in a manner closely
analogous to that considered here.

\section{Discussion}

We present a single-integral solution to the angular correlation
function, by considering the special case of a uniform galaxy
distribution. This provides a useful alternative to the
traditional Limber and thin-layer approximations, and offers a
substantial improvement in accuracy when evaluating the appearance
and location of the acoustic peak.

This top-hat approximation can be utilised to derive an analytic
estimate for the impact of photometric redshift errors on BAO
surveys. In agreement with Zhan \& Knox
\cite{2006ApJ...644..663Z}, we find the value of $\sigma_z$ ought
to be known to better than $\sim 10^{-3}$ if the acoustic peak is
to provide significant and robust constraints on the dark energy
equation of state, and this is particularly apparent for lower
redshifts, $z<1$. It also seems likely that the anticipated
departure from Gaussianity of this error distribution must be well
understood.

We have not addressed the complexities associated with galaxy
bias, since the same projection process applies regardless of the
original form of the spatial correlation function. However, within
the context of a photometric redshift survey a well-defined model
of $\xi(r)$ will be required in order to reverse the smoothing
process.

Whilst future projects such as LSST and SNAP will look to higher
redshifts, it is the near-term projects such as Pan-STARRS and DES
which, despite their higher tolerance for statistical errors, may
be more susceptible to the difficulties in mapping the
low-redshift behaviour of $\sigma_z(z)$.

\noindent{\bf Acknowledgements}\\
FS was funded by an STFC rolling grant, and is grateful for the
generous welfare support from the University of Edinburgh. PS
acknowledges support by the European DUEL Research-Training
Network (MRTN-CT-2006-036133).

\bibliography{M:/Routines/dis}

\begin{thebibliography}{14}
\expandafter\ifx\csname natexlab\endcsname\relax\def\natexlab#1{#1}\fi
\expandafter\ifx\csname bibnamefont\endcsname\relax
  \def\bibnamefont#1{#1}\fi
\expandafter\ifx\csname bibfnamefont\endcsname\relax
  \def\bibfnamefont#1{#1}\fi
\expandafter\ifx\csname citenamefont\endcsname\relax
  \def\citenamefont#1{#1}\fi
\expandafter\ifx\csname url\endcsname\relax
  \def\url#1{\texttt{#1}}\fi
\expandafter\ifx\csname urlprefix\endcsname\relax\def\urlprefix{URL }\fi
\providecommand{\bibinfo}[2]{#2}
\providecommand{\eprint}[2][]{\url{#2}}

\bibitem[{\citenamefont{{Blake} and {Glazebrook}}(2003)}]{2003ApJ...594..665B}
\bibinfo{author}{\bibfnamefont{C.}~\bibnamefont{{Blake}}} \bibnamefont{and}
  \bibinfo{author}{\bibfnamefont{K.}~\bibnamefont{{Glazebrook}}},
  \bibinfo{journal}{\apj} \textbf{\bibinfo{volume}{594}}, \bibinfo{pages}{665}
  (\bibinfo{year}{2003}).

\bibitem[{\citenamefont{{Seo} and {Eisenstein}}(2003)}]{2003ApJ...598..720S}
\bibinfo{author}{\bibfnamefont{H.-J.} \bibnamefont{{Seo}}} \bibnamefont{and}
  \bibinfo{author}{\bibfnamefont{D.~J.} \bibnamefont{{Eisenstein}}},
  \bibinfo{journal}{\apj} \textbf{\bibinfo{volume}{598}}, \bibinfo{pages}{720}
  (\bibinfo{year}{2003}).

\bibitem[{\citenamefont{{Eisenstein} et~al.}(2005)\citenamefont{{Eisenstein},
  {Zehavi}, {Hogg}, {Scoccimarro}, {Blanton}, {Nichol}, {Scranton}, {Seo},
  {Tegmark}, {Zheng} et~al.}}]{2005ApJ...633..560E}
\bibinfo{author}{\bibfnamefont{D.~J.} \bibnamefont{{Eisenstein}}},
  \bibinfo{author}{\bibfnamefont{I.}~\bibnamefont{{Zehavi}}},
  \bibinfo{author}{\bibfnamefont{D.~W.} \bibnamefont{{Hogg}}},
  \bibinfo{author}{\bibfnamefont{R.}~\bibnamefont{{Scoccimarro}}},
  \bibinfo{author}{\bibfnamefont{M.~R.} \bibnamefont{{Blanton}}},
  \bibinfo{author}{\bibfnamefont{R.~C.} \bibnamefont{{Nichol}}},
  \bibinfo{author}{\bibfnamefont{R.}~\bibnamefont{{Scranton}}},
  \bibinfo{author}{\bibfnamefont{H.-J.} \bibnamefont{{Seo}}},
  \bibinfo{author}{\bibfnamefont{M.}~\bibnamefont{{Tegmark}}},
  \bibinfo{author}{\bibfnamefont{Z.}~\bibnamefont{{Zheng}}},
  \bibnamefont{et~al.}, \bibinfo{journal}{\apj} \textbf{\bibinfo{volume}{633}},
  \bibinfo{pages}{560} (\bibinfo{year}{2005}).

\bibitem[{\citenamefont{{Cole} et~al.}(2005)\citenamefont{{Cole}, {Percival},
  {Peacock}, {Norberg}, {Baugh}, {Frenk}, {Baldry}, {Bland-Hawthorn},
  {Bridges}, {Cannon} et~al.}}]{2005MNRAS.362..505C}
\bibinfo{author}{\bibfnamefont{S.}~\bibnamefont{{Cole}}},
  \bibinfo{author}{\bibfnamefont{W.~J.} \bibnamefont{{Percival}}},
  \bibinfo{author}{\bibfnamefont{J.~A.} \bibnamefont{{Peacock}}},
  \bibinfo{author}{\bibfnamefont{P.}~\bibnamefont{{Norberg}}},
  \bibinfo{author}{\bibfnamefont{C.~M.} \bibnamefont{{Baugh}}},
  \bibinfo{author}{\bibfnamefont{C.~S.} \bibnamefont{{Frenk}}},
  \bibinfo{author}{\bibfnamefont{I.}~\bibnamefont{{Baldry}}},
  \bibinfo{author}{\bibfnamefont{J.}~\bibnamefont{{Bland-Hawthorn}}},
  \bibinfo{author}{\bibfnamefont{T.}~\bibnamefont{{Bridges}}},
  \bibinfo{author}{\bibfnamefont{R.}~\bibnamefont{{Cannon}}},
  \bibnamefont{et~al.}, \bibinfo{journal}{\mnras}
  \textbf{\bibinfo{volume}{362}}, \bibinfo{pages}{505} (\bibinfo{year}{2005}).

\bibitem[{\citenamefont{{Blake} and {Bridle}}(2005)}]{2005MNRAS.363.1329B}
\bibinfo{author}{\bibfnamefont{C.}~\bibnamefont{{Blake}}} \bibnamefont{and}
  \bibinfo{author}{\bibfnamefont{S.}~\bibnamefont{{Bridle}}},
  \bibinfo{journal}{\mnras} \textbf{\bibinfo{volume}{363}},
  \bibinfo{pages}{1329} (\bibinfo{year}{2005}),
  \eprint{arXiv:astro-ph/0411713}.

\bibitem[{\citenamefont{{Cai} et~al.}(2008)\citenamefont{{Cai}, {Angulo},
  {Baugh}, {Cole}, {Frenk}, and {Jenkins}}}]{2008arXiv0810.2300C}
\bibinfo{author}{\bibfnamefont{Y.-C.} \bibnamefont{{Cai}}},
  \bibinfo{author}{\bibfnamefont{R.~E.} \bibnamefont{{Angulo}}},
  \bibinfo{author}{\bibfnamefont{C.~M.} \bibnamefont{{Baugh}}},
  \bibinfo{author}{\bibfnamefont{S.}~\bibnamefont{{Cole}}},
  \bibinfo{author}{\bibfnamefont{C.~S.} \bibnamefont{{Frenk}}},
  \bibnamefont{and}
  \bibinfo{author}{\bibfnamefont{A.}~\bibnamefont{{Jenkins}}},
  \bibinfo{journal}{ArXiv e-prints}  (\bibinfo{year}{2008}),
  \eprint{0810.2300}.

\bibitem[{\citenamefont{{Simon}}(2007)}]{2006astro.ph..9165S}
\bibinfo{author}{\bibfnamefont{P.}~\bibnamefont{{Simon}}},
  \bibinfo{journal}{\aap} \textbf{\bibinfo{volume}{473}}, \bibinfo{pages}{711}
  (\bibinfo{year}{2007}), \eprint{arXiv:astro-ph/0609165}.

\bibitem[{\citenamefont{{LoVerde} and {Afshordi}}(2008)}]{2008arXiv0809.5112L}
\bibinfo{author}{\bibfnamefont{M.}~\bibnamefont{{LoVerde}}} \bibnamefont{and}
  \bibinfo{author}{\bibfnamefont{N.}~\bibnamefont{{Afshordi}}},
  \bibinfo{journal}{ArXiv e-prints}  (\bibinfo{year}{2008}),
  \eprint{0809.5112}.

\bibitem[{\citenamefont{{Lewis} et~al.}(2000)\citenamefont{{Lewis},
  {Challinor}, and {Lasenby}}}]{2000ApJ...538..473L}
\bibinfo{author}{\bibfnamefont{A.}~\bibnamefont{{Lewis}}},
  \bibinfo{author}{\bibfnamefont{A.}~\bibnamefont{{Challinor}}},
  \bibnamefont{and}
  \bibinfo{author}{\bibfnamefont{A.}~\bibnamefont{{Lasenby}}},
  \bibinfo{journal}{\apj} \textbf{\bibinfo{volume}{538}}, \bibinfo{pages}{473}
  (\bibinfo{year}{2000}), \eprint{arXiv:astro-ph/9911177}.

\bibitem[{\citenamefont{{Smith} et~al.}(2003)\citenamefont{{Smith}, {Peacock},
  {Jenkins}, {White}, {Frenk}, {Pearce}, {Thomas}, {Efstathiou}, and
  {Couchman}}}]{2003MNRAS.341.1311S}
\bibinfo{author}{\bibfnamefont{R.~E.} \bibnamefont{{Smith}}},
  \bibinfo{author}{\bibfnamefont{J.~A.} \bibnamefont{{Peacock}}},
  \bibinfo{author}{\bibfnamefont{A.}~\bibnamefont{{Jenkins}}},
  \bibinfo{author}{\bibfnamefont{S.~D.~M.} \bibnamefont{{White}}},
  \bibinfo{author}{\bibfnamefont{C.~S.} \bibnamefont{{Frenk}}},
  \bibinfo{author}{\bibfnamefont{F.~R.} \bibnamefont{{Pearce}}},
  \bibinfo{author}{\bibfnamefont{P.~A.} \bibnamefont{{Thomas}}},
  \bibinfo{author}{\bibfnamefont{G.}~\bibnamefont{{Efstathiou}}},
  \bibnamefont{and} \bibinfo{author}{\bibfnamefont{H.~M.~P.}
  \bibnamefont{{Couchman}}}, \bibinfo{journal}{\mnras}
  \textbf{\bibinfo{volume}{341}}, \bibinfo{pages}{1311} (\bibinfo{year}{2003}),
  \eprint{arXiv:astro-ph/0207664}.

\bibitem[{\citenamefont{{Newman}}(2008)}]{2008ApJ...684...88N}
\bibinfo{author}{\bibfnamefont{J.~A.} \bibnamefont{{Newman}}},
  \bibinfo{journal}{\apj} \textbf{\bibinfo{volume}{684}}, \bibinfo{pages}{88}
  (\bibinfo{year}{2008}), \eprint{0805.1409}.

\bibitem[{\citenamefont{{Zhan} and {Knox}}(2006)}]{2006ApJ...644..663Z}
\bibinfo{author}{\bibfnamefont{H.}~\bibnamefont{{Zhan}}} \bibnamefont{and}
  \bibinfo{author}{\bibfnamefont{L.}~\bibnamefont{{Knox}}},
  \bibinfo{journal}{\apj} \textbf{\bibinfo{volume}{644}}, \bibinfo{pages}{663}
  (\bibinfo{year}{2006}), \eprint{arXiv:astro-ph/0509260}.

\bibitem[{\citenamefont{{Ma} et~al.}(2006)\citenamefont{{Ma}, {Hu}, and
  {Huterer}}}]{2006ApJ...636...21M}
\bibinfo{author}{\bibfnamefont{Z.}~\bibnamefont{{Ma}}},
  \bibinfo{author}{\bibfnamefont{W.}~\bibnamefont{{Hu}}}, \bibnamefont{and}
  \bibinfo{author}{\bibfnamefont{D.}~\bibnamefont{{Huterer}}},
  \bibinfo{journal}{\apj} \textbf{\bibinfo{volume}{636}}, \bibinfo{pages}{21}
  (\bibinfo{year}{2006}), \eprint{arXiv:astro-ph/0506614}.

\bibitem[{\citenamefont{{Ma} and {Bernstein}}(2008)}]{2008ApJ...682...39M}
\bibinfo{author}{\bibfnamefont{Z.}~\bibnamefont{{Ma}}} \bibnamefont{and}
  \bibinfo{author}{\bibfnamefont{G.}~\bibnamefont{{Bernstein}}},
  \bibinfo{journal}{\apj} \textbf{\bibinfo{volume}{682}}, \bibinfo{pages}{39}
  (\bibinfo{year}{2008}), \eprint{0712.1562}.

\end{thebibliography}

\end{document}